\documentclass{article}
\usepackage{spconf,amsmath,graphicx,url,times, hyperref}
\usepackage[hyphenbreaks]{breakurl}
\usepackage{color}
\usepackage{booktabs}
\usepackage{setspace, enumitem}

\title{Extending Audio Masked Autoencoders toward Audio Restoration}

\name{%
\begin{tabular}{@{}c@{}}
Zhi Zhong$^{1}$, Hao Shi$^{2}$, Masato Hirano$^{1}$, Kazuki Shimada$^{3}$, Kazuya Tateishi$^{1}$\\
Takashi Shibuya$^{3}$, Shusuke Takahashi$^{1}$, Yuki Mitsufuji$^{1,3}$
\end{tabular}}
\address{$^{1}$ Sony Group Corporation, Japan \qquad $^{2}$ Kyoto University, Japan \qquad $^{3}$ Sony Research, Japan
}

\begin{document}
\ninept
\maketitle

\begin{sloppy} 

\begin{abstract}
\vspace{-1mm}
Audio classification and restoration are among major downstream tasks in audio signal processing. However, restoration derives less of a benefit from pretrained models compared to the overwhelming success of pretrained models in classification tasks. Due to such unbalanced benefits, there has been rising interest in how to improve the performance of pretrained models for restoration tasks, \textit{e.g.}, speech enhancement (SE). Previous works have shown that the features extracted by pretrained audio encoders are effective for SE tasks, but these speech-specialized encoder-only models usually require extra decoders to become compatible with SE, and involve complicated pretraining procedures or complex data augmentation. Therefore, in pursuit of a universal audio model, the audio masked autoencoder (MAE) whose backbone is the autoencoder of Vision Transformers (ViT-AE), is extended from audio classification to SE, a representative restoration task with well-established evaluation standards. ViT-AE learns to restore masked audio signal via a mel-to-mel mapping during pretraining, which is similar to restoration tasks like SE. We propose variations of ViT-AE for a better SE performance, where the mel-to-mel variations yield high scores in non-intrusive metrics and the STFT-oriented variation is effective at intrusive metrics such as PESQ. Different variations can be used in accordance with the scenarios. Comprehensive evaluations reveal that MAE pretraining is beneficial to SE tasks and help the ViT-AE to better generalize to out-of-domain distortions. We further found that large-scale noisy data of general audio sources, rather than clean speech, is sufficiently effective for pretraining.
\end{abstract}
\vspace{-2.5mm}
\section{Introduction}
\label{sec:intro}
\vspace{-2mm}
Various methods have been proposed to pretrain models that can extract generally useful features from audio, with the expectation that these features will benefit multiple downstream tasks. Audio classification and restoration are two major downstream tasks in audio signal processing, but restoration tasks have typically derived less benefit from pretrained models. For example, application of pretrained models to audio classification tasks such as audio scene classification, keyword spotting, and music instrument classification has shown great success \cite{koutini2022patchout,ntt2022msmmae,peking2022maskspec,meta2022audiomae,zhong2023attention}. However, audio restoration tasks are often tackled by training models from scratch, as in the case of speech enhancement (SE) or bandwidth extension (BWE) \cite{liu2022voicefixer, dolby2022universe}. Due to the unbalanced benefit, there is rising interest in how to improve the performance of pretrained models for restoration tasks. In this work, SE is chosen as a representative of numerous restoration tasks due to its well-established objective metrics and test data \cite{dolby2022universe}. While \cite{kong2021speech} tried to apply a pretrained audio classifier to create paired data for the conditional source separation task, the separation model did not work well in SE tasks.

Some methods exploit models that are pretrained via self-supervised learning (SSL) for SE tasks. Inspired by BERT \cite{devlin2018bert}, various SSL models for speech have been proposed, such as HuBERT \cite{hsu2021hubert} and WavLM \cite{chen2022wavlm}. These models are pretrained with the mask prediction task, in which the masked tokens are predicted from visible tokens. While the features extracted by these models can be used for SE tasks \cite{song2022exploreWavLM, huang2022investSSL}, there are some drawbacks. First, these models have no decoder, which makes it necessary to select an extra decoder to become compatible with SE tasks \cite{song2022exploreWavLM, huang2022investSSL}. Moreover, the models have been pretrained with only speech data, which limits their extension to tasks other than speech. In addition, the pretraining involves complicated procedures such as the estimation of pseudo labels \cite{hsu2021hubert,chen2022wavlm} or complex data augmentations such as speech overlapping or noise simulation \cite{chen2022wavlm,song2022exploreWavLM}.

A framework that is directly compatible with restoration tasks like SE, easy to pretrain and less demanding in data augmentation has yet to be explored. Audio masked autoencoders (MAE) \cite{ntt2022msmmae,peking2022maskspec,meta2022audiomae} are SSL methods that have been reported to have state-of-the-art (SOTA) performance in audio classification tasks \cite{ntt2022msmmae,peking2022maskspec,meta2022audiomae}. Audio MAE can be simply pretrained by ground truth labels, and is able to reduce resource consumption by not inputting masked tokens to the model \cite{he2022mae}. The backbone of MAE is the autoencoder of Vision Transformers \cite{dosovitskiy2020vit} (ViT-AE), which learns to restore audio signal via a mel-to-mel mapping during pretraining. However, excluding qualitative experiments on packet loss concealment \cite{meta2022audiomae}, audio MAE has not been considered for SE or other restoration tasks.

To address the above issues, first, we \textbf{extend audio MAE to speech enhancement} and explore the framework's potential for solving multiple tasks such as audio classification and SE. The ViT-AE backbone negates the needs for pseudo labels and naturally learns a mel-to-mel mapping that is compatible with SE tasks during MAE pretraining. Second, we \textbf{propose variations of ViT-AE} to improve the performance on SE tasks, where the mel-to-mel variations have advantages in terms of non-intrusive metrics and the STFT-oriented variation is effective for standard intrusive metrics such as PESQ. The aforementioned variations can be utilized in accordance with the scenarios. Finally, we carried out comprehensive evaluations and ablation studies, in which we found that \textbf{large-scale noisy data of general audio sources}, rather than clean speech, is \textbf{sufficiently effective} for pretraining. Pretraining with general audio data also allows us to further explore tasks other than speech (\textit{e.g.}, music) in the future. We encourage our readers to visit the webpage for audio samples and an appendix about music BWE \footnote[1]{\url{https://zzaudio.github.io/Demo\_Extend\_AudioMAE\_toward\_Restoration/demo\_page.html}}.
\vspace{-3mm}
\section{Related Work}
\label{sec:related_work}
\vspace{-2mm}
\begin{figure}[tb]
    \includegraphics[width= \linewidth]{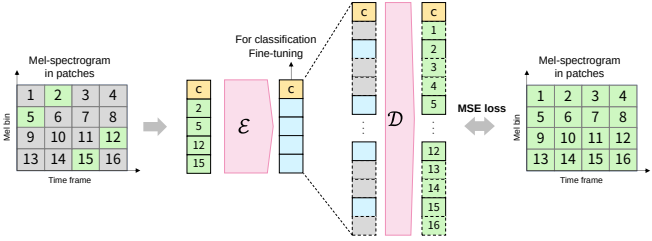} %
    \centering
    \vspace{-7mm}
    \caption{Audio MAE pretraining. Each patch includes 16-by-16 time-mel bins. Masked patches are deleted from the input sequence and dummy tokens are used in the decoder to reconstruct the input. The cls (``c'' in orange) token is used for classification tasks.}
    \label{fig:mae}
    \vspace{-5mm}
\end{figure}

A number of prior studies have explored the application of speech-specialized BERT-like SSL models to SE tasks. Song \textit{et al.} combined noisy short-time Fourier transform (STFT) magnitude with features extracted by WavLM-base+ as the input to the backend speech enhancers and found the extracted features useful for SE tasks \cite{song2022exploreWavLM}. Huang \textit{et al.} compare features extracted by pretrained encoders  with standard STFT and log-mel spectrograms in \cite{huang2022investSSL}, and found that pretraining models on clean speech may result in domain mismatch in SE tasks, as these models have to process noisy rather than clean speech in such cases. Hung \textit{et al.} \cite{hung2022boosting} showed that the SE performance of pretrained encoders could be boosted by combining multiple techniques with a careful fine-tuning strategy. However, as mentioned in Sec.\ref{sec:intro}, the above speech-specialized pretrained models are not naturally compatible with SE tasks, which also suffer from complicated pretraining procedures, complex data augmentation and heavy computation during pretraining.

Audio MAE pretraining is often conducted with log-mel spectrogram, hence ViT-AE naturally learns the mel-to-mel mapping. Huang \textit{et al.} \cite{meta2022audiomae} qualitatively confirmed that the pretrained ViT-AE can achieve packet loss concealment without fine-tuning, but did not perform any in-depth study for such restoration tasks. For speech enhancement via mel-to-mel mapping, it is common to use a pretrained neural vocoder to convert the processed log-mel spectrogram back into raw waves \cite{liu2022voicefixer,haoyu2021se_voco}.%
\vspace{-2.5mm}
\section{Approach}
\label{sec:approach}
\vspace{-2mm}
\subsection{Audio MAE for Audio Classification}
\label{ssec:mae}
\vspace{-1mm}
Audio MAE methods \cite{ntt2022msmmae,peking2022maskspec, meta2022audiomae} originate from image MAE \cite{he2022mae} and typically take the log-mel spectrogram as input. Unlike conventional Transformer methods in speech processing, which consider the time axis in raw waves or spectrograms as the direction of input sequences \cite{hsu2021hubert,chen2022wavlm,chen2021conformer}, ViT divides log-mel into patches (each patch contains 16-by-16 time-mel bins following \cite{peking2022maskspec,meta2022audiomae,he2022mae,dosovitskiy2020vit}), and conducts sequential modelling in accordance with the patch's position in the original log-mel spectrogram, as shown in Fig. \ref{fig:mae}. A large portion (around 75\%) of the input patches are masked to create the mask prediction task. In contrast to \cite{hsu2021hubert,chen2022wavlm}, the ViT encoder is followed by a decoder in audio MAE, and hence the backbone is called ViT-AE. Thanks to the decoder, pseudo labels are not needed for pretraining. The decoder also allows masked patches to be exempted from the input, thereby greatly reducing the computational cost. We insert a cls token as the feature vector for downstream classification tasks, as it will automatically aggregate the features from other tokens when fine-tuned \cite{koutini2022patchout,he2022mae}.
\vspace{-2.5mm}
\subsection{ViT-AE Framework for Multiple Tasks}
\label{ssec:vit-ae}
\vspace{-1mm}
The vanilla ViT-AE learns mel-to-mel mapping via mean square error (MSE) loss in Fig. \ref{fig:mae}. Although the vanilla ViT-AE architechture is directly applicable to downstream restoration tasks, we will show in Sec.\ref{ssec:res_se} that the speech enhancement quality of vanilla ViT-AE is not sufficient. Some works try to improve the reconstruction quality of MAE-based frameworks by modifying the loss function in pretraining \cite{song2022exploreWavLM} or modifying the model structure \cite{li2022mage}. Unfortunately, modification during pretraining will largely affect the classification performance of MAE-based methods \cite{li2022mage}. We therefore introduce new training losses and variations of ViT-AE, which we use only for fine-tuning ViT-AE on SE tasks. This should bring no change to pretraining procedures hence no harm to other tasks such as classification, while simultaneously offering apparent benefits to SE.

\begin{figure}[tb]
    \includegraphics[width= \linewidth]{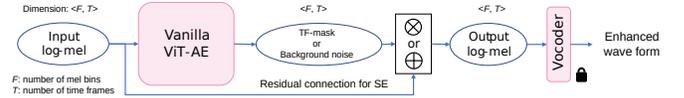} %
    \centering
    \vspace{-6.5mm}
    \caption{Mel-to-mel ViT-AE with residual connection.}
    \label{fig:vit-ae}
    \vspace{-2mm}
\end{figure}
\begin{figure}[tb]
    \includegraphics[width= \linewidth]{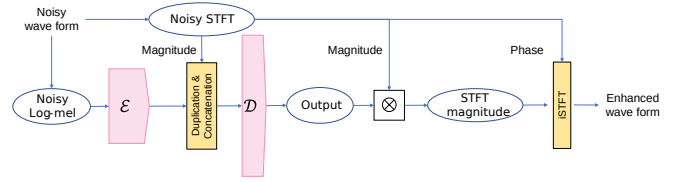} %
    \centering
    \vspace{-6.5mm}
    \caption{ViT-AE-iSTFT, where the decoder takes the concatenation of STFT magnitude and ViT encoder features as input.}
    \label{fig:vit-ae-iSTFT}
    \vspace{-6mm}
\end{figure}

\noindent\textbf{Mel-to-mel ViT-AE with residual connection}. We introduce a residual connection outside the vanilla ViT-AE to better enhance the log-mel spectrogram, which results in two variations shown in Fig. \ref{fig:vit-ae}: additive and multiplicative ViT-AE, both of which originate from the domain knowledge of audio signal processing. Modelling the input noisy speech as the addition of clean signal and background noises, the additive ViT-AE learns to enhance signals by subtracting background noises from the noisy input. The multiplicative ViT-AE estimates time-frequency (TF) masks for mel-spectrograms. Masks are limited to the range of [0, 1] by a sigmoid function to indicate how much energy the clean speech occupies in each bin of the noisy mel-spectrogram. Masks are then multiplied with noisy input. We do not directly apply masks to the log-mel, as log-scale power compression results in negative values, making the physical meaning of a non-negative mask unclear. Log-mel spectrograms are converted into raw waves via a pretrained frozen vocoder (Fig. \ref{fig:vit-ae}).

To better capture details of spectrograms from both log- and linear-scale, the proposed loss for SE fine-tuning is the summation of MSE losses in both scales of mel-spectrogram, \textit{i.e.},
\vspace{-1.2mm}
\begin{equation}
    \label{eq:melloss}
    L_\mathrm{SE} = \mathrm{MSE}(\hat{Y}, Y) + \lambda\cdot\mathrm{MSE}(\mathrm{exp}(\hat{Y}), \mathrm{exp}(Y)),
    \vspace{-1.8mm}
\end{equation}
where $Y$ is the ground truth log-mel spectrogram of clean speech, $\hat{Y}$ is the estimated one by ViT-AE, and the exponential function recovers the linear mel-spectrogram from log-scale power compression. The coefficient of $\lambda = 100$ is determined empirically to balance losses from both scales.

\noindent\textbf{STFT-oriented ViT-AE}. Mel-to-mel ViT-AE's objective scores (such as PESQ) are limited and upper bounded by the vocoder \cite{liu2022voicefixer}. To improve the objective scores of ViT-AE, we propose a variation to generate enhanced waves via inverse STFT (iSTFT) (ViT-AE-iSTFT in Fig. \ref{fig:vit-ae-iSTFT}). Inspired by \cite{song2022exploreWavLM}, the ViT decoder (different from the one used in pretraining) takes the concatenation of noisy STFT magnitude and ViT encoder features as the input, in which the ViT encoder features are expected to provide auxiliary information to improve the SE performance of the decoder. Following the common practice in \cite{song2022exploreWavLM,huang2022investSSL,hung2022boosting}, STFT magnitude is enhanced by an STFT mask (similar to the mel-to-mel multiplicative ViT-AE). To fine-tune this variation, unlike equation (\ref{eq:melloss}), L1 loss is empirically found more suitable for the linear scale STFT.

While usages similar to Fig. \ref{fig:vit-ae-iSTFT} have been explored for HuBERT- and WavLM-based methods \cite{song2022exploreWavLM,huang2022investSSL,hung2022boosting}, it is non-trivial to evaluate such usages for ViT-AE. In contrast to HuBERT or WavLM, the ViT-AE takes time-mel patches as input, which means the direction of input sequences is not aligned with the time axis in STFT spectrograms (Sec.\ref{ssec:mae}). To concatenate the ViT encoder features with STFT, we have to extend the length of ViT encoder features by duplicating them twice under our current setting, which makes it even more difficult to extract information about the time axis. Moreover, there is a domain mismatch for ViT-AE, because it is pretrained to learn mel-to-mel mapping, not the STFT mapping. Nonetheless, in Sec.\ref{ssec:res_se} we will show that this ViT-AE-iSTFT achieves high objective scores, which means the ViT encoder can provide valuable auxiliary information under the above limitations.
\vspace{-3mm}
\section{Experiments}
\label{sec:exp}
\vspace{-1.5mm}
We evaluate the ViT-AE and MAE pretraining on both audio classification and speech enhancement tasks.
\vspace{-3.25mm}
\subsection{MAE Pretraining}
\label{ssec:pretrain}
\vspace{-1.25mm}
We carried out MAE pretraining with 75\% mask ratio for the ViT-AE on either AudioSet (noisy but general) or LibriTTS (clean and speech-specialized) to examine the influence of the pretraining data. AudioSet \cite{google2017audio} contains around 2 million 10-second audio segments taken from YouTube and annotated with 527 diverse classes. AudioSet has been widely used in general audio representation learning \cite{koutini2022patchout,ntt2022msmmae,peking2022maskspec,meta2022audiomae, kong2020panns}. The unbalanced set of AudioSet is used for pretraining. LibriTTS \cite{zen2019libritts} is a large-scale corpus of English audiobooks presented at sentence break. We merge all three train subsets and two dev subsets of the LibriTTS for pretraining. 

The ViT and the masking strategy are implemented based on the open-source project in \cite{koutini2022patchout, rw2019timm}. Hyperparameters of the ViT-AE are based on \cite{ntt2022msmmae}, where the ViT encoder is a ViT-base model, containing 12 Transformer blocks, whose attention layer has 768 dimensions, and the number of attention heads is 12. The decoder is a smaller ViT, with four layers of 384 dimensions and eight attention heads. The dimension of the feedforward networks is four times the attention layer dimension across the whole ViT-AE.

Following \cite{ntt2022msmmae}, audio data is cropped or padded to 5-second chunks (448 frames) during pretraining. ViT-AE is pretrained for 60 epochs with the batch size of 128 and learning rate (LR) of 1e-4. The LR linearly increases to 1e-4 in the first five epochs (warmup), and then decreases to 1e-6 by the cosine annealing scheduler \cite{peking2022maskspec, meta2022audiomae}. Instead of audio overlapping or noise simulation, we apply only random gain (--6 dB to +3 dB) and cyclic cropping \cite{koutini2022patchout} as data augmentation. To alleviate the affect of different audio loudness, all input log-mel spectrograms are normalized by the train set mean and standard deviation that are collected in advance \cite{meta2022audiomae}. The AdamW optimizer is used with a weight decay of 1e-4 \cite{meta2022audiomae}.

The 80-bin log-mel is generated to be compatible with the vocoder's settings \cite{kong2020hifigan}: raw waves at a sampling rate of 22.05 kHz are processed by STFT with a 1024-point Hann window and 256-point hop size, followed by a mel filterbank. Throughout experiments, audio files are resampled to 22.05 kHz if needed.
\vspace{-3mm}
\subsection{Downstream: Audio Classification}
\label{ssec:classification}
\vspace{-1.25mm}
The cls token of the pretrained ViT encoder is followed by a single linear layer during the fine-tuning for classification tasks. The speech classification accuracy of ViT-AE is evaluated by Speech Command V2 (SPCv2) \cite{warden2018spcv2}, a dataset that presents a 35-class single-label speech command recognition task. When ViT-AE is pretrained with AudioSet (as in \cite{ntt2022msmmae,peking2022maskspec,meta2022audiomae}), its ability in general audio classification is measured by the 527-class multi-label task on AudioSet-2M (summation of unbalanced and balanced subsets) and AudioSet-20k (the balance subset alone).
\vspace{-2.75mm}
\subsection{Downstream: Speech Enhancement}
\label{ssec:se}
\vspace{-1.25mm}
We fine-tune ViT-AE and its variations on the standard Valentini's dataset \cite{valentini2016VBD}, whose test set has 824 noisy speeches without reverberation. Generalization ability is measured by a subset of the DAPS dataset \cite{mysore2014can} under a zero-shot condition. By playing back pre-recorded clean speeches in noisy and reverberate environments, distorted speeches from 20 speakers under 12 scenarios are created by consumer-grade devices. We constructed the out-of-domain subset using all recordings of the last female and male speakers. 

During the finetuning, the batch size is 16, the LR increases to 1e-4 within the 10 epochs of warmup and decreases to 1e-6 by 90 epochs of cosine annealing. To convert log-mel spectrograms into raw waves, we use the HiFi-GAN vocoder \cite{kong2020hifigan}. The mel-to-mel ViT-AE keeps the same settings as in pretraining, while the ViT-AE-iSTFT extends the decoder's attention layer dimensions to 512 to predict the one-side STFT (rather than mel-spectrogram) masks.
\begin{table}[tb]
    \vspace{-2.25mm}
    \caption{Results of audio classification. AudioSet: Mean average precision (\%). SPCv2: Top-1 accuracy (\%). For SPCv2, either AudioSet or LibriTTS pretraining yields the same score for our model.}
    \label{tab:res_classification}
    \centering
    \resizebox{7 cm}{!}{
    \begin{tabular}{cccc}
    \toprule
    {}&{ViT-AE/Audio MAE (ours)}&{PANN \cite{kong2020panns}}&{MaskSpec \cite{peking2022maskspec}}\\
    \midrule
    {AS-2M} & {43.4} & {43.1} & {47.1}\\
    {AS-20k} & {31.7} & {27.8} & {32.3}\\
    \hline
    {SPCv2} & {97.8} & {--} & {97.7}\\
    \bottomrule
    \end{tabular}}
\vspace{-5mm}
\end{table}
\begin{table}[b]
    \vspace{-8mm}
    \caption{SE: From-scratch vs. pretrained (Bold: Top-2 scores)}
    \label{tab:res_scratch_vs_pretrain}
    \centering
    \resizebox{7.5cm}{!}{
    \begin{tabular}{ccccccc}
    \toprule
    {Pretrain}&{Method}&{PESQ}&{Csig}&{Cbak}&{Covrl}&{NISQA}\\
    \midrule
    {}&{Noisy}&{1.97}&{3.35}&{\textbf{2.44}}&{2.63}&{3.34}\\
    \hline
    {}&{Vanilla ViT-AE}&{2.22}&{3.80}&{2.36}&{3.00}&{3.98}\\
    {Scratch}&{Additive (Fig. \ref{fig:vit-ae})}&{2.34}&{3.88}&{2.38}&{3.10}&{4.18}\\
    {}&{Multiplicative (Fig. \ref{fig:vit-ae})}&{2.35}&{3.90}&{2.38}&{3.11}&{4.18}\\
    \hline
    {}&{Vanilla ViT-AE}&{2.23}&{3.83}&{2.37}&{3.02}&{4.05}\\
    {LibriTTS}&{Additive}&{2.39}&{3.95}&{2.41}&{3.16}&{4.27}\\
    {}&{Multiplicative}&{\textbf{2.42}}&{\textbf{3.97}}&{\textbf{2.42}}&{\textbf{3.18}}&{\textbf{4.29}}\\
    \hline
    {}&{Vanilla ViT-AE}&{2.23}&{3.83}&{2.37}&{3.01}&{4.06}\\
    {AudioSet}&{Additive}&{2.40}&{3.96}&{2.41}&{3.17}&{4.25}\\
    {}&{Multiplicative}&{\textbf{2.42}}&{\textbf{3.98}}&{\textbf{2.42}}&{\textbf{3.19}}&{\textbf{4.29}}\\
    \hline
    {}&{Vocoder Oracle}&{3.01}&{4.68}&{2.81}&{3.87}&{4.57}\\
    {}&{Clean}&{4.50}&{5.00}&{5.00}&{5.00}&{4.62}\\
    \bottomrule
    \end{tabular}}
\end{table}
\vspace{-3mm}
\section{Evaluation}
\label{sec:eval}
\vspace{-2mm}
\subsection{Audio Classification}
\label{ssec:res_classification}
\vspace{-2mm}
The results presented in Tab. \ref{tab:res_classification} demonstrate that we have successfully reproduced audio MAE. Our AudioSet model outperforms PANN in general audio classification tasks, showing its ability to tasks other than speech. Both the AudioSet and LibriTTS models achieve 97.8\% accuracy in SPCv2, outperforming MaskSpec \cite{peking2022maskspec} (an audio MAE model) in the speech classification task. MaskSpec  obtained higher scores at AudioSet, which can be explained by differences in hyperparameters (we followed \cite{ntt2022msmmae}, but no AudioSet scores were reported there) and loss in the training data (downloadable segments from YouTube differ by region, and decrease over time).

\vspace{-2.75mm}
\subsection{Speech Enhancement}
\label{ssec:res_se}
\vspace{-1.25mm}
We evaluate the importance of pretraining by standard intrusive metrics (PESQ, Csig, Cbak and Covrl \cite{metrics}) and a non-intrusive metric called NISQA \cite{mittag2021nisqa} in Valentini's dataset. First, the additive and multiplicative ViT-AE stably outperform the vanilla version in every pretraining condition as shown in Tab. \ref{tab:res_scratch_vs_pretrain}. Next, \textbf{pretrained ViT-AE outperforms the from-scratch} one for all variations of ViT-AE in all metrics, though the difference brought by different pretraining data was trivial. Last but not least, the additive and multiplicative variations reflect more benefits of pretraining compared to the vanilla one, revealing that the \textbf{proposed variations are important} to unleash the power of MAE pretraining in SE tasks.
\begin{table}[htb]
    \vspace{-5.5mm}
    \caption{SE: ViT-AE-iSTFT vs Mel-to-mel ViT-AE (Bold: Top 1)}
    \label{tab:res_stft_vs_mel}
    \centering
    \resizebox{7.5 cm}{!}{
    \begin{tabular}{ccccccc}
    \toprule
    {Pretrain}&{Method}&{PESQ}&{Csig}&{Cbak}&{Covrl}&{NISQA}\\
    \midrule
    {}&{noisy}&{1.97}&{3.35}&{2.44}&{2.63}&{3.34}\\
    \hline
    {Scratch}&{ViT decoder with STFT input}&{2.71}&{3.96}&{2.59}&{3.33}&{3.95}\\
    {}&{ViT-AE-iSTFT (Fig. \ref{fig:vit-ae-iSTFT})}&{2.75}&{3.99}&{2.61}&{3.37}&{4.00}\\
    \hline
    {LibriTTS}&{ViT-AE-iSTFT}&{2.80}&{4.06}&{2.64}&{3.43}&{4.03}\\
    \hline
    {AudioSet}&{ViT-AE-iSTFT}&{\textbf{2.85}}&{\textbf{4.12}}&{\textbf{2.67}}&{\textbf{3.48}}&{4.05}\\
    {}&{Multiplicative ViT-AE (Fig. \ref{fig:vit-ae})}&{2.42}&{3.98}&{2.42}&{3.19}&{\textbf{4.29}}\\
    \hline
    {}&{Clean}&{4.50}&{5.00}&{5.00}&{5.00}&{4.62}\\
    \bottomrule
    \end{tabular}}
    \vspace{-2.5mm}
\end{table}

The multiplicative ViT-AE is compared with ViT-AE-iSTFT in Tab. \ref{tab:res_stft_vs_mel}. First, ViT-AE-iSTFT successfully improves objective scores of the decoder-only baseline, which implies that the ViT encoder can provide auxiliary information to improve the SE performance of the decoder despite the limitations mentioned in Sec.\ref{ssec:vit-ae}. The AudioSet pretraining is slightly more beneficial than LibriTTS for ViT-AE-iSTFT. Next, our goal to further improve the objective scores of mel-to-mel variations by ViT-AE-iSTFT is partly realized, as ViT-AE-iSTFT outperforms the multiplicative one in \textbf{every intrusive metric} especially in PESQ (2.85 vs 2.42). Meanwhile, the merit of using mel-to-mel mapping with a vocoder is revealed by the higher NISQA score of the multiplicative version. NISQA helps us to further clarify  the
differences among the proposed ViT-AE variations, because standard intrusive metrics may contrast with human perception when evaluating vocoder-based methods \cite{liu2022voicefixer}.  We therefore conclude that mel-to-mel ViT-AE should be utilized for cases requiring perceptual quality (NISQA) and that ViT-AE-iSTFT is better for scenarios requiring high intrusive scores (PESQ).

\begin{table}[b]
    \vspace{-8mm}
    \caption{SE: Comparison with existing methods (Bold: Ours). $^{*}$UNIVERSE: Unofficial implementation by the authors.} 
    \label{tab:res_comparison}
    \centering
    \resizebox{7.5cm}{!}{
    \begin{tabular}{ccccccc}
    \toprule
    {Comment}&{Method}&{PESQ}&{Csig}&{Cbak}&{Covrl}&{NISQA}\\
    \midrule
    {-}&{Noisy}&{1.97}&{3.35}&{2.44}&{2.63}&{3.34}\\
    \hline
    {Pretrained \&}&{\textbf{Multiplicative ViT-AE}}&{2.42}&{3.98}&{2.42}&{3.19}&{4.29}\\
    {GAN (vocoder)}&{Vocoder Oracle}&{3.01}&{4.68}&{2.81}&{3.87}&{4.57}\\
    \hline
    {GAN}&{SEGAN \cite{pascual2017segan}}&{2.16}&{3.48}&{2.84}&{2.80}&{-}\\
    {}&{Voicefixer \cite{liu2022voicefixer}}&{2.43}&{-}&{-}&{-}&{-}\\
    \hline
    {}&{\textbf{ViT-AE-iSTFT}}&{2.85}&{4.12}&{2.67}&{3.48}&{4.05}\\
    {Pretrained \&}&{PANN + UNet \cite{kong2021speech}}&{2.28}&{2.43}&{2.96}&{2.30}&{-}\\
    {Discriminative}&{Huang \textit{et al.} \cite{huang2022investSSL}}&{2.68}&{-}&{-}&{-}&{-}\\
    {}&{Hung \textit{et al.} \cite{hung2022boosting}}&{3.16}&{4.50}&{3.57}&{3.86}&{-}\\
    \hline
    {}&{UNIVERSE$^{*}$ \cite{dolby2022universe}}&{2.90}&{4.03}&{3.11}&{3.46}&{4.61}\\
    {Diffusion}&{SGMSE+ \cite{richter2022sgmse+}}&{2.94}&{4.25}&{3.40}&{3.61}&{4.56}\\
    {}&{GP-Unified \cite{anonymous2023gpunified}}&{2.95}&{4.18}&{3.44}&{3.57}&{4.61}\\
    \hline
    {Discriminative}&{DCUNet \cite{choi2019dcunet}}&{3.13}&{4.24}&{4.00}&{3.69}&{-}\\
    \hline
    {-}&{Clean}&{4.50}&{5.00}&{5.00}&{5.00}&{4.62}\\
    \bottomrule
    \end{tabular}}
\end{table}
We compare the ViT-AE pretrained on AudioSet with other methods in Tab. \ref{tab:res_comparison}. The multiplicative version obtained a similar PESQ score as Voicefixer\cite{liu2022voicefixer} which also uses a vocoder, and a higher PESQ score than SEGAN \cite{pascual2017segan}. Although ViT-AE-iSTFT has lower metrics than SOTA discriminative models like \cite{choi2019dcunet}, it achieves comparable intrusive metrics to SOTA diffusion models such as UNIVERSE, SGMSE+ and its variation GP-Unified \cite{dolby2022universe,richter2022sgmse+,anonymous2023gpunified}, except in Cbak. These diffusion models  achieved NISQA scores comparable to the oracle results of vocoder, revealing the limited potential of vocoder-based methods. Note that the methods compared above are \textbf{not capable} of classification.

For methods compatible with classification, we exempt results from HuBERT-large and WavLM-large for fair comparison. Variations of ViT-AE outperform PANN + UNet \cite{kong2021speech} and Huang \textit{et al.} \cite{huang2022investSSL} in all metrics except Cbak. Hung \textit{et al.} \cite{hung2022boosting} utilized multiple techniques with a careful fine-tuning strategy in which the model is partially frozen, and as a result, their method even outperforms SOTA models in some metrics. However, speech-specialized pretrained models used in \cite{hung2022boosting} are not reported to be beneficial for non-speech tasks such as AudioSet classification. We leave the improvement of ViT-AE's objective scores to future work.

Finally, we evaluate the generalization ability by applying models trained or fine-tuned on Valentini's dataset to the unseen DAPS dataset (zero-shot condition), and report the mean scores of all 12 scenarios in Tab. \ref{tab:res_generalization}. Contrary to Tab. \ref{tab:res_comparison}, we found the pretrained multiplicative ViT-AE outperforming UNIVERSE, indicating that our system generalized better to unseen scenarios. While SGMSE+ is reported to have SOTA generalization ability \cite{richter2022sgmse+}, the multiplicative ViT-AE pretrained on AudioSet achieved similar Csig and Covrl scores, even if ViT-AE is not an SE-specialized method. As for the pretrained mel-to-mel ViT-AE variations, all outperformed the from-scratch ones. Moreover, the AudioSet pretraining outperformed the LibriTTS one in all variations except the NISQA score for ViT-AE-iSTFT, which implies that the \textbf{diverse audio sources and recording environments} in AudioSet have been beneficial for generalization. ViT-AE-iSTFT didn't consistently benefit from pretraining,  which could have been caused by limitations mentioned in Sec. \ref{ssec:vit-ae} and is worth investigation in the future.
\begin{table}[tb]
    \vspace{-2mm}
    \caption{SE: Out-of-domain Generalization (Bold: Top-2 scores).}
    \label{tab:res_generalization}
    \centering
    \resizebox{7.5cm}{!}{
    \begin{tabular}{cccccc}
    \toprule
    {Comment}&{Method}&{PESQ}&{Csig}&{Covrl}&{NISQA}\\
    \midrule
    {}&{noisy}&{1.47}&{2.00}&{1.60}&{2.79}\\
    \hline
    {}&{UNIVERSE$^{*}$ (unofficial) \cite{dolby2022universe}}&{1.51}&{2.19}&{1.76}&{3.79}\\
    {}&{SGMSE+ \cite{richter2022sgmse+}}&{\textbf{1.97}}&{2.81}&{\textbf{2.33}}&{\textbf{4.44}}\\
    {}&{GP-Unified \cite{anonymous2023gpunified}}&{\textbf{1.88}}&{2.81}&{2.27}&{\textbf{4.32}}\\
    \hline
    {}&{Vanilla ViT-AE}&{1.67}&{2.77}&{2.14}&{3.14}\\
    {Scratch}&{Additive}&{1.65}&{2.63}&{2.05}&{3.58}\\
    {}&{Multiplicative}&{1.64}&{2.68}&{2.08}&{3.49}\\
    {}&{ViT decoder with STFT input}&{1.79}&{2.44}&{2.01}&{2.86}\\
    {}&{ViT-AE-iSTFT}&{1.81}&{2.54}&{2.07}&{2.92}\\
    \hline
    {}&{Vanilla ViT-AE}&{1.72}&{2.81}&{2.19}&{3.11}\\
    {LibriTTS}&{Additive}&{1.73}&{2.84}&{2.21}&{3.77}\\
    {}&{Multiplicative}&{1.74}&{2.87}&{2.23}&{3.80}\\
    {}&{ViT-AE-iSTFT}&{1.81}&{2.49}&{2.05}&{2.95}\\
    \hline
    {}&{Vanilla ViT-AE}&{1.76}&{2.86}&{2.24}&{3.15}\\
    {AudioSet}&{Additive}&{1.78}&{\textbf{2.93}}&{2.29}&{3.79}\\
    {}&{Multiplicative}&{1.79}&{\textbf{2.95}}&{\textbf{2.30}}&{3.82}\\
    {}&{ViT-AE-iSTFT}&{1.83}&{2.59}&{2.11}&{2.90}\\
    \hline
    {}&{Vocoder Oracle}&{2.99}&{4.66}&{3.84}&{4.72}\\
    {}&{Clean}&{4.50}&{5.00}&{5.00}&{4.78}\\
    \bottomrule
    \end{tabular}}
    \vspace{-6mm}
\end{table}
%

\vspace{-2.5mm}
\section{Conclusion}
\label{sec:conclusion}
\vspace{-1.5mm}
In pursuit of universal audio models, this paper extended audio MAE from audio classification to speech enhancement (SE). SE was chosen from among many restoration tasks due to its widely accepted evaluation benchmark. The output features of pretrained audio encoders are known to be effective for SE tasks, but prior speech-specialized encoder-only models usually require extra decoders to become compatible with SE tasks, and involve complicated pretraining procedures or complex data augmentation. To address these issues, audio MAE, \textit{i.e.}, ViT-AE with MAE pretraining, was explored. Audio MAE simplifies the pretraining procedures, and enables the ViT-AE to naturally learn mel-to-mel mapping that is compatible with SE tasks. We proposed variations of ViT-AE to improve the SE performance and unleash the power of pretraining. Comprehensive evaluations and ablation studies demonstrate that MAE pretraining is beneficial for SE tasks and help the ViT-AE to better generalize to out-of-domain test data. We further found that large-scale noisy data of general audio sources, rather than clean speech, are sufficiently effective for pretraining. We leave the improvement of objective metrics as our future work, as well as the in-depth exploration to a broader range of restoration tasks with various sound sources . We hope our findings here pave the way to the continued research on universal audio models.

\vfill\pagebreak
\begin{spacing}{0.925}
\bibliographystyle{IEEEtran}
\bibliography{refs23}
\end{spacing}

\end{sloppy}
\end{document}